\documentclass[journal,comsoc]{IEEEtran}

\usepackage[T1]{fontenc}

\ifCLASSINFOpdf
  
\else

\fi

\usepackage{amsmath}

\usepackage[cmintegrals]{newtxmath}

\usepackage{pifont} 		    
\usepackage{threeparttable}     
\usepackage{multirow}        	
\usepackage{tabularx}       	
\usepackage{graphicx}           
\usepackage{booktabs}			
\usepackage{hyperref}           
\usepackage{color}              

\begin{document}

\title{A Secure Authentication Framework to Guarantee the Traceability of Avatars in Metaverse}

\author{

Kedi~Yang, 
Zhenyong~Zhang$^*$,
Youliang~Tian,~\IEEEmembership{Member,~IEEE,}
and Jianfeng~Ma,~\IEEEmembership{Member,~IEEE}

\thanks{Manuscript received XX, 2022; revised XX, 2022; accepted
XX, 2022. Date of publicationXX, 2022; date of current version XX, 2022.  This paper is supported by the National Key Research and Development Program of China under Grant No.2021YFB3101100; Key Program of the National Natural Science Union Foundation of China under Grant No.U1836205; Project of High-level Innovative Talents of Guizhou Province under Grant No. [2020]6008; Science and Technology Program of Guiyang under Grant No.[2021]1-5, No.[2022]2-4; Science and Technology Program of Guizhou Province under Grant No. [2020]5017, No. [2022]065. Guizhou Provincial Postgraduate Research Fund under Grant No.YJSKYJJ[2021]029. The associate editor coordinating the review of this manuscript and approving it for publication was XXX. \emph{ ( Corresponding author: Youliang Tian. )} }
\thanks{Kedi Yang, Zhenyong Zhang, and Youliang Tian are with the State Key Laboratory of Public Big Data, College of Computer Science and Technology, 
Guizhou University, Guiyang 550025, China, and also with the Institute of Cryptography \& Data Security, GuiZhou University, Guiyang 550025, China  (e-mail: kdyang.gz@gmail.com; zyzhangnew@gmail.com; youliangtian@163.com).}
\thanks{Jianfeng Ma is with the School of Cyber Engineering, Xidian University,
Xi’an 710126, China, and also with the State Key Laboratory of Public Big
Data, College of Computer Science and Technology, Guizhou University,
Guiyang 550025, China (e-mail: jfma@mail.xidian.edu.cn).

$^*$ means equal contribution with the first author.}
}

\markboth{Submitted to IEEE Trans, 2022}
{Shell \MakeLowercase{\textit{et al.}}: Bare Demo of IEEEtran.cls for IEEE Communications Society Journals}

\maketitle

\begin{abstract}
Metaverse is a vast virtual environment parallel to the physical world in which users enjoy a variety of services acting as an avatar. To build a secure living habitat, it’s vital to ensure the virtual-physical traceability that tracking a malicious player in the physical world via his avatars in virtual space. In this paper, we propose a two-factor authentication framework based on chameleon signature and biometric-based authentication. First, aiming at disguise in virtual space, we propose a chameleon collision signature algorithm to achieve the verifiability of the avatar's virtual identity. Second, facing at impersonation in physical world, we construct an avatar's identity model based on the player's biometric template and the chameleon key to realize the verifiability of the avatar's physical identity. Finally,  we design two decentralized authentication protocols based on the avatar's identity model to ensure the consistency of the avatar's virtual and physical identities.  Security analysis  indicates that the proposed authentication framework guarantees the consistency and traceability of avatar's identity. Simulation experiments show that the framework not only completes the  decentralized authentication between avatars but also achieves the virtual-physical tracking.
\end{abstract}


\IEEEpeerreviewmaketitle

\section{Introduction}

\IEEEPARstart{M}{etaverse}, combination of the prefix “meta” implying transcending with the stem “verse” of the universe \cite{Lee2021All}, means a new type of Internet application and social form beyond the physical world. The word first appeared in the science fiction novel, \emph{Snow Crash}, written by Neal Stephenson in 1992, which described a vast virtual environment parallel to the physical world, where people communicate and work through digital avatars.

The most representative prototype of metaverse is the virtual platform \emph{Second Life} \footnote{ https://secondlife.com } released by Linden Research in 2003. The platform constructs a virtual world that is highly similar to reality in which players can freely socialize, trade, and build facilities via their digital role. IBM once purchased a piece of land and built its own sales center in the complete virtual ecosystem. As a highly digitized world, metaverse is devoted to build an environment that satisfies immersive interaction and virtual-physical coexistence \cite{Cheng2022Will},  which breaks through the limitations of the physical world and enables people to perform unimaginable works. For example, in the medical field, the digital representation of a patient is projected into a virtual consultation room, in which experts from around the world discuss treatment options face-to-face without leaving their office. In education, virtual molecules and atoms are cast around students to perceive the abstract microscopic world. In the social field, friends in different places are projected into the same virtual environment allowing them to communicate and shop like in the physical world.

Metaverse will become the second living space of human beings coexistence with the physical world \cite{Duan2021Social}.  At present,  leading companies in various countries have turned their attention to the metaverse  \cite{Kraus2022Facebook} \cite{Yang2022Survey},  where the dimension is infinite but the ecosystem is finite.  Baidu, the largest Internet company in China, has published a immersive interactive environment, \emph{Xirang} \footnote{https://vr.baidu.com/product/xirang },  based on Virtual Reality (VR) and Artificial Intelligence (AI). The environment consists of infinitely connected virtual spaces, each of them is a unique digital metropolis. MATE (formerly Facebook), the most popular social platform in America, has released an open VR social environment \emph{Horizon Worlds} that enables each player to create his own community and meet strangers from all over the world for leisure and entertainment. However, something as disturbing as the real world is happening in this emerging environment. In the public beta of \emph{Horizon Worlds}, a female tester reported that her avatar was sexually harassed by other players. Soon after, it was reported that a researcher from SumOfUs had suffered similar harassment in the virtual cyberspace, which was worse than the previous. It can be seen from the above events that the avatar's safety is being threatened, which seriously hinders the further development of metaverse \cite{Wang2022Privacy}. Therefore, there is an urgent need to establish a traceable authentication mechanism that tracks a malicious avatar to its player.

Realizing such a traceable mechanism is non-trivial. One of the challenges is that we are unable to track malicious players via the avatars' appearances because different players may create avatars with the same shape \cite{Yu2021AvatarEmbody}. Another challenge is that it is impossible to track the malicious player directly through the device in which the malicious player may somehow obtain devices authenticated from legitimate users to impersonate the player. The final challenge is that mutual authentication between avatars may impose a massive overload to the system. The reason is that metaverse breaks through the physical limitations in terms of geography, allowing millions of users to communicate in a
 zone.

In this paper, we aim to address the above challenges. For the first challenge, the player can generate a signature when creating an avatar, thereby forming non-repudiation information for the avatar's virtual identity. However, the traditional signature algorithm \cite{Thyagarajan2020Signatures} is cumbersome, which results in that players not only have to perform a lot of calculations but also need to store a large number of signature parameters for later tracing. Therefore, it is essential to construct an efficient signature algorithm to ensure the traceability of the avatar's virtual identity. Chameleon signature is a one-to-many signature mechanism that signs multiple plaintexts with one signed message. Based on this feature, the players are able to sign multiple avatars without changing the original signed message, which significantly improves computational efficiency and reduces storage costs. However, the traditional chameleon signature consists of two parts, the chameleon hash and the common signature algorithm, which leads to the players having to hold multiple key pairs. To this end, we modify the chameleon hash function \cite{Khalili2019Efficient} and propose a chameleon collision signature algorithm, which enables players to sign avatars with only a pair of keys.

For the second challenge, a natural idea is that the user as a verifier throws a random challenge to the avatar, while the avatar as a prover  submits its player's biological sample in response. Based on these physical identity parameters, the verifier can check the avatar's physical identity. However, the biometric feature is difficult to avoid replaying attacks \cite{Liang2020Continuous} and thus a malicious player may deny that the violation wasn't performed by him. To solve this problem, we build an avatar's identity model based on the player's biometric template and chameleon key. In the process of authentication, it requires the player to embed a watermarking into the captured biological sample and generate a check parameter utilizing his chameleon private key to guarantee the verifiability of the avatar's physical identity.

For the third challenge, we assume that there is an authentication mechanism that allows avatars to complete dynamic identity verification without involving a trusted third party, which ensures the consistency of the avatar's virtual and physical identities in a decentralized manner. Unfortunately, current mutual authentication schemes primarily focus on one-time authentication between devices \cite{Luo2020OcuLock} and have not yet considered dynamic authentication between avatars. To this end, we designed two immersive decentralized authentication protocols based on the avatar's identity model, including the one-party authentication protocol and the two-parties authentication protocol. By these protocols, players can implement dynamic authentication as well as track malicious avatars.

To sum up, we construct a decentralized and traceable authentication framework for avatars based on chameleon signature and biometrics. The contributions are as follows:

\begin{itemize}
\item We introduce the notion of virtual-physical tracking for avatars authentication  systems and define the security requirement  of consistency for avatar's virtual and physical identities.
\item We propose a chameleon collision signature algorithm to achieve the verifiability of the avatar's virtual identity, which effectively generates chameleon collisions to form signatures using only one private key.
\item We construct an avatar's identity model based on the biometric template and the chameleon public key to realize the verifiability of the avatar's physical identity.
\item We design two sets of decentralized avatar authentication protocols based on the avatar's identity model to guarantee the consistency of the avatar's virtual and physical identities. 
\item We build an avatar authentication system based on blockchain and iris recognition method, which realizes the virtual-physical tracking to avatars.
\end{itemize}

The rest of the paper is organized as follows. We introduce the secure authentication framework in the next section. The verifiable avatar's identity is presented in Section \ref{sec:AvatarIdent}. Section \ref{sec:AuthProto} designs the decentralized avatar authentication protocol. The security analysis and the performance evaluation are given in Section \ref{sec:SecuAnaly} and Section \ref{sec:PerfEval},  respectively. We review related works in Section \ref{sec:RelaWork} and ﬁnally conclude the paper in Section \ref{sec:Conclu}.

\section{Secure Authentication Framework} \label{sec:AuthFrawe}
\begin{figure}[htbp]
\begin{center}
    \includegraphics[width=0.47\textwidth]{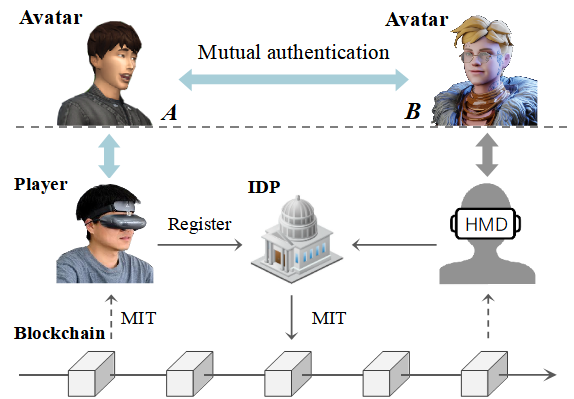}
    \caption{\small{System model of the security authentication framework.}}
    \label{fig:SysModel}
 \end{center}
\end{figure}

To ensure the virtual-physical traceability tracking a virtual avatar to its physical player, we combine knowledge-based and biometric-based methods to build a two-factor  authentication framework. Here, we introduce the system model, security threats, and design goals of the proposed authentication framework shown in Fig. \ref{fig:SysModel}.

\subsection{System Model}
\begin{itemize}
    \item \emph{Avatar:} It is the virtual representation of a physical player. Based on this feature, we regard an avatar as the unity of virtual identity and physical identity, which means that a valid avatar must satisfy the consistency of virtual and physical identities.
    \item \emph{Player:} It is a specific manipulator of an avatar in the physical world. To ensure traceability, players need to submit their chameleon key and biological samples to complete registration and obtain a verifiable metaverse identity token (MIT), which enables the player to create an avatar that satisfies the virtual-physical traceability.
    \item \emph{Identity provider (IDP):} As a trusted organization in the physical world, it audits players' real identities and generates the corresponding MIT. During the tracking process, it discloses the violator's real identity through the MIT and the related parameters submitted by whistleblower.
    \item \emph{Blockchain:} It stores public information about the avatars' identity.

\end{itemize}

\subsection{Security Threats}
A malicious player may disguise his avatar as the one of target player's  to deceive others, may get device authenticated by legitimate player to manipulate the corresponding avatar, and may even replay outdated information collected  from interactions to target player. Thus, we assume that the malicious player can mount the following attacks:

\begin{itemize}
    \item \emph{Disguise:} In the virtual metaverse, the attacker as a malicious avatar disguises his appearance that looks the same as the target avatar to deceive the interactor.
    \item \emph{Impersonation:} In the physical world, the attacker as a malicious manipulator gets a device authenticated by legitimate player and impersonates the player to manipulate the corresponding avatar.
    \item \emph{Replay:} Both in the metaverse and the physical world, the attacker collects the outdated identity parameters associated with a honest avatar  and submits them to the interactor by which claims to be the target avatar.
\end{itemize}

\subsection{Design Goals}
To realize the secure interaction for avatars in the metaverse, the proposed framework should achieve the following goals.

\begin{itemize}
    \item \emph{Traceability:} The traceability refers to tracking an avatar in the virtual metaverse  to the corresponding manipulator in the physical world, we call it virtual-physical traceability.  It requires that a malicious avatar should be traced back to its manipulator through the avatar's identity parameters retained in the authentication process,  while a honest player should not be framed by fake identity parameters about his avatar. 
    \item \emph{Consistency:} To achieve virtual-physical traceability, the avatar's virtual and physical identities must be consistent. Therefore,  we should check the validity of the avatar's virtual identity to prevent disguise, check the avatar's physical identity to avoid impersonation, and check the freshness of the submitted identity parameters to prevent replaying.
    \item \emph{Decentralized:} Authentication between avatars can be done without the involvement of a trusted third party.
    \item \emph{Immersive:} During the authentication process, players do not need to perform specific operations, ensuring the immersive experience for players.
    \item \emph{Privacy:} The player's physical identity is not disclosed during authentication.
\end{itemize}

\section{Verifiable Avatar's Identity} \label{sec:AvatarIdent}
An avatar is a unity of the virtual identity and the physical identity. In this section, we first propose a chameleon collision signature algorithm to ensure the verifiability of the avatar's virtual identity, which prevents malicious players from mounting disguise. Then, we design an avatar's identity model to ensure the verifiablity of avatar's physical identity, which prevents impersonation.

\subsection{Chameleon Collision Signature}\label{sec:ChamSign}
The methods of traditional chameleon signature  involve signing on the chameleon hash and forging collisions, resulting in the signer having to hold two private keys, one for signing and the other for generating collisions. In addition, the methods introduce zero-knowledge proofs to ensure the security of chameleon hash , which seriously reduces the efficiency of the algorithm \cite{Khalili2019Efficient} \cite{Camenisch2017Trapdoors}.

Since a collision is forged by the private key, the new collision can be treated as a chameleon signature. Based on this idea, we propose an efficient chameleon collision signature, which reduces the cost of holding keys and eliminates the zero-knowledge proofs while ensuring security by modifying the chameleon hash \cite{Khalili2019Efficient}. The proposed algorithm consists of the following six parts, namely, $Setup$, $KeyGen$, $Hash$, $Check$, $Sign$, and $Verify$.

\begin{itemize}

\item $Setup(\lambda)\rightarrow Parm$. Let $\lambda$ be a security parameter in the chameleon collision signature system. $\mathbb{G}, \mathbb{G}_T$ are multiplicative cyclic groups of prime order $q\geq 2^\lambda $, where $g $ is a generator of $\mathbb{G}$. A pairing  $\hat{e}:\mathbb{G}\times\mathbb{G}\rightarrow \mathbb{G}_T$ is an eﬃciently computable bilinear map, which satisfies $\hat{e}(g^a,g^b)=\hat{e}(g,g)^{ab}$  for all $a,b\in \mathbb{Z}_q$.  The system selects the global anti-collision hash function $H_\mathbb{G}:\{0,1\}^*\rightarrow\mathbb{G}$ , which maps bit strings of arbitrary length to corresponding elements in $\mathbb{G}$. Finally, the algorithm  $Setup$ publishes the system parameters $Parm=\{\mathbb{G},\mathbb{G}_T,g_,q,\hat{e},H_\mathbb{G}\}$.
 
\item $KeyGen(Parm)\rightarrow (pk,sk) $.  The key generation algorithm takes the system parameter $Parm$ as input. The algorithm picks a randomness $x\stackrel{R}{\leftarrow}  \mathbb{Z}_q$ as the private key $sk$ and calculates $y=g^{(1/x)}\in \mathbb{G}$ as the public key $pk$. It outputs the following public-private key pair 	
$$sk = x, \; pk = y. $$

\item $Hash(pk,M) \rightarrow (h,R)$. The chameleon hash generation algorithm takes as input the public key $pk=y$ and the message $M$. The algorithm sets $m=H_\mathbb{G}(M)$ and picks a randomness $r\stackrel{R}{\leftarrow} \mathbb{Z}_q$ . It outputs the chameleon hash $h$ and the corresponding check parameter $R$ as 
\begin{equation} \label{equ:Hash}
    h=m \cdot y^r,  \quad  			R=g^r.  \nonumber
\end{equation}

\item $Check(pk,h,M,R)\rightarrow b$. The chameleon hash check algorithm takes as input the public key $pk=y$,  the chameleon hash $h$, the message $M$, and the check parameter $R$.  The algorithm sets $m=H_\mathbb{G}(M)$ and checks the compatibility of $(y,h,m,R)$.  It outputs $b=1$ if
\begin{equation} \label{Check}
    \hat{e}(h/m,g)=\hat{e}(R,y) . \nonumber
\end{equation}

\item $Sign(sk,h,M,R,M^\prime)\rightarrow R^\prime$. To sign a message $M^\prime$, the chameleon sign  algorithm takes as input the chameleon private key $ sk = x $ and the chameleon triple $(h,M,R)$. The algorithm sets $m^\prime=H_\mathbb{G}(M^\prime)$ and outputs the check parameter $R^\prime$ as 
\begin{equation} 
    R^\prime=(h/m^\prime)^{x} . \nonumber
\end{equation}Where $(M,R)$ and $(M^\prime, R^\prime)$ are called a chameleon collision pair with respect to $h$.

\item $Verify(pk,h,M,R,M^\prime,R^\prime) \rightarrow b $.  To verify the signature of collision pair $(M,R)$ and $(M^\prime, R^\prime)$ , the algorithm checks the compatibility of $(pk,h,M,R)$ and  $(pk,h,M^\prime,R^\prime)$. It outputs $b=1$ if
\begin{equation} 
    Check(pk,h,M,R) = Check(pk,h,M^\prime,R^\prime) = 1 .\nonumber
\end{equation}
\end{itemize}

\subsection{Avatar's Identity Model}
In this part, we design an avatar's identity model based on the chameleon signature and the biometrics, including the  metaverse identity token (MIT), avatar's virtual identity (VID), and the avatar's physical identity (PID). As shown in Fig.\ref{fig:OnePartyProcess}, the player's chameleon public key $pk=y \in MIT $ is linked to the avatar $A$'s virtual identity $VID=(M_a,R_a)$. The player's biological template $T \in MIT$ is related to the avatar's physical identity  $PID=(M_a^\prime, R_a^\prime)$ , which ensures the verifiability of avatar's physical identity. 

\textbf{Metaverse Identity Token (MIT):}
The metaverse identity token is a bridge between the virtual space and the physical world. To obtain this token $MIT$, the user provides $ID$, $M$, $T$, and $y$ to IDP, then the IDP generates a signed $MIT = (T, y, h, M, R )$ after reviewing the above information. Among them, the $ID$ is the player's real identity, $M$ is the anonymous identity, $T$ is the biometric template such as the iris, $y$ is the chameleon public key taken from the key pair $(x,y) \leftarrow KeyGen( \cdot )$, $h$ is  the chameleon hash generated by $(h, R) \leftarrow Hash( y, M )$, and $R$ is the check parameter of $M$. Ultimately, the IDP publishes $MIT$ on the blockchain to ensure the public verifiability of the avatar's identity and records $(ID, MIT)$ in a secure database to guarantee the traceability of the player's real identity.

\textbf{Virtual Identity (VID):}
The avatar's virtual identity (VID) is the player's public identity in a metaverse region, which appears as a visible avatar. Before entering the metaverse, the player need to create the $VID = ( M_a, R_a )$ based on the $MIT$ and his chameleon private key $x$, where $M_a$ is the avatar's identity information, $R_a \leftarrow Sign( sk, h, M, R,  M_a )$ is the corresponding check parameter.

\textbf{Physical Identity (PID):}
The avatar's physical identity (PID) is the player's biometric information in the physical world, which presents as a processed biometric feature. During metaverse interactions, an avatar provides its $PID=(M_a^\prime, R_a^\prime)$ to the verifier, where $M_a^\prime$ is the player's biometric feature, $R_a^\prime \leftarrow Sign( sk, h, M, R,  M_a^\prime )$ is the corresponding check parameter.

\section{Decentralized Avatar Authentication Protocol}\label{sec:AuthProto}
In this section, we design two decentralized authentication protocols based on the avatar's identity mode. In the one-party authentication protocol, the verifier throws challenges to the prover at any time to achieve real-time authentication. For the two-party authentication, the verifier and the prover can not only complete mutual authentication but also establish a session key to facilitate secure communication.

\subsection{One-party Authentication Protocol}
The  one-party authentication protocol implements dynamic authentication based on a challenge-response mechanism to ensure the consistency of avatar's virtual and physical identities. As shown in Fig.
\ref{fig:OnePartyProcess}: 1) avatar $A$ as a prover claims that his identity is valid; 2) avatar $B$ as a verifier checks the validity of $A$'s virtual identity and throws a random challenge to $A$ to confirm whether $A$'s physical identity matches its virtual identity; 3)  avatar $A$'s manipulator provides his biometric feature and corresponding check parameters as a response; 4) avatar $B$ checks the validity of the parameters to determine the consistency of avatar's virtual and physical identities. The detailed processes are as follows:

\begin{figure}[htbp]
\begin{center}
    \includegraphics[width=0.47\textwidth]{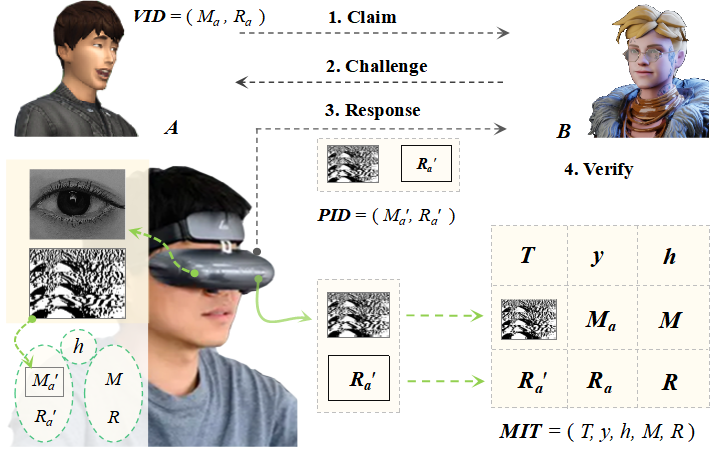}
    \caption{\small{The interaction process of one-party authentication protocol.}}
    \label{fig:OnePartyProcess}
 \end{center}
 \end{figure}

\textbf{Claim:} 
In this phase, avatar $A$ submits its metaverse identity token $MIT = ( T, y, h, M, R )$ and virtual identity information $VID=(M_a,R_a)$ to avatar $B$ to initiate an identity claim.

\begin{figure}[htbp]
\begin{center}
    \includegraphics[width=0.47\textwidth]{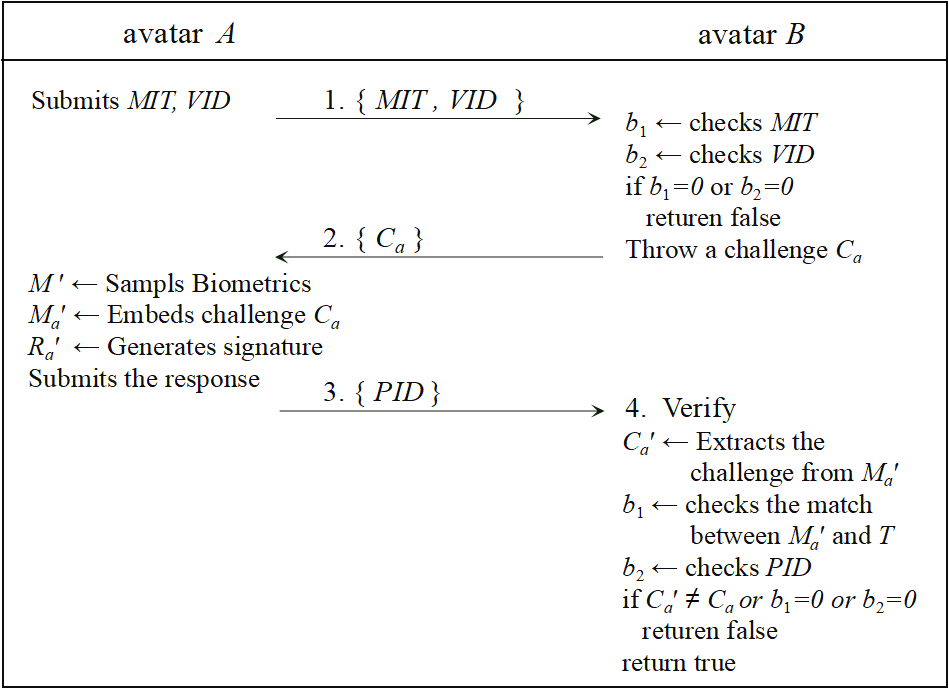}
    \caption{\small{One-party authentication protocol.}}
    \label{fig:OnePartyProtocol}
 \end{center}
 \end{figure}

\textbf{Challenge:}
After avatar $B$ completing the check on $MIT$ and $VID$, he throws a random challenge $C_a$ to avatar $A$. As shown in Fig.\ref{fig:OnePartyProtocol}, if the IDP's signature on $MIT$ is validity and the collision  $(M_a,R_a)=VID$ with respect to $(M,R) \in MIT$ is validity according to $Verify ( y, h, M, R,$ $M_a, R_a )$ in \ref{sec:ChamSign},  $A$'s virtual identity is validity.

\textbf{Response:}
In the response phase,  avatar $A$'s manipulator submits his biometrics embedded with challenge information to prove the validity of $A$'s physical identity. The specific steps are as follows: (i) the manipulator randomly samples his biometric information and processes it locally to form a biometric feature $M^\prime$; (ii) the manipulator embeds the challenge $C_a$ as watermarking into $M^\prime$ to form a watermarked biometric feature $M_a^\prime$, which provides conditions for anti-replay attacks; (iii) the manipulator generates the check parameter $R_a^\prime \leftarrow Sign(sk,h,M,R,M_a^\prime)$ using his chameleon private key $sk$ according to \ref{sec:ChamSign}. Finally, $A$ submits $PID=(M_a^\prime, R_a^\prime)$ to avatar $B$ as the response corresponding to $C_a$.

\textbf{Verify:}
To verify whether avatar $A$'s physical identity is consistent with his virtual identity, the avatar $B$  needs to perform the following three steps: (i) extracts the challenge information $C_a^\prime$ from $M_a^\prime$ and checks $C_a^\prime=C_a$ to ensure whether $M_a^{\prime}$ is a playback;  (ii) checks the match between the biometric feature $M_c^\prime$ and the biological template $T \in MIT$ to ensure the validity of $A$'s physical identity; (iii) checks the match between the collision pairs $(M_a^\prime, R_a^\prime)=PID$ and $(M,R) \in MIT$ to ensure the consistency of avatar’s virtual and physical identities. After the above steps are passed, the verifier can determine that $A$'s physical identity is consistent with its virtual identity.

\subsection{Two-party Authentication Protocol}\label{sec:TwoPartAuth}
In this part, we designed a two-party authentication protocol to realize the decentralized mutual authentication between avatars. As shown in Fig. \ref{fig:TwoPartyProtocl}, the designed protocol adds a session key negotiation based on the one-party protocol to achieve secure communication between avatar $A$ and avatar $B$. The detailed processes are in the following three phases:

\begin{figure}[htbp]
\begin{center}
    \includegraphics[width=0.47\textwidth]{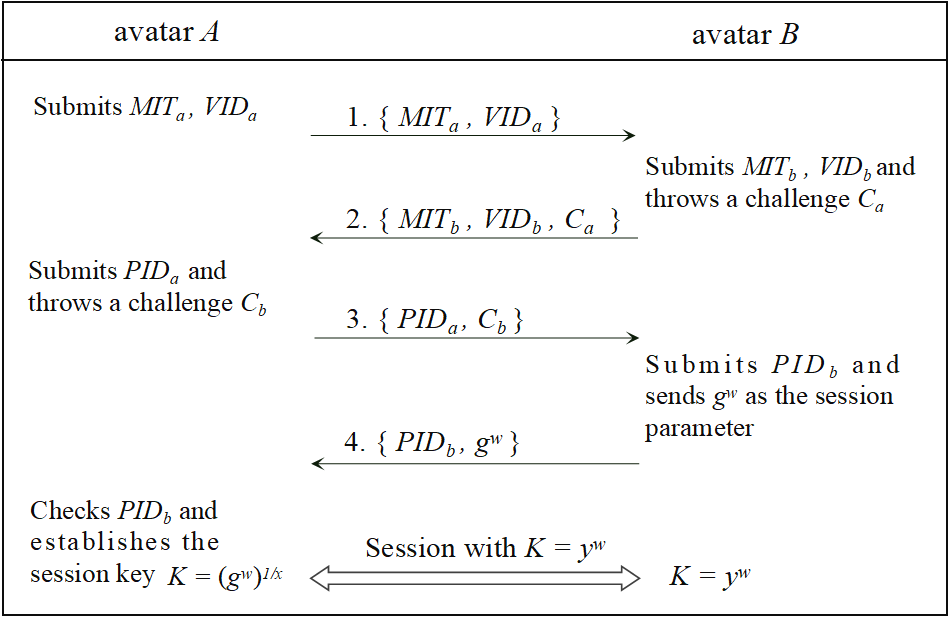}
    \caption{\small{Two-parties authentication protocol.}}
    \label{fig:TwoPartyProtocl}
 \end{center}
 \end{figure}

\textbf{Round 1:} 
In this phase, first avatar $A$ submits an identity claim to avatar $B$, then $B$ submits a claim of $B$'s identity and  a challenge to $A$. The specific steps are as follows: (i) $A$ submits  $B$ with $MIT_a$ and $VID_a$ to initiate a claim about $A$'s identity; (ii) upon $B$ checking the validity of $A$'s virtual identity based on $MIT_a$ and $VID_a$, $B$ sends $MIT_b, VID_b$ and $C_a$ to $A$, where $MIT_b$ and $VID_b$ is the claim of $B$'s identity, $C_a$ is a random challenge to $A$'s physical identity.

\textbf{Round 2:}
 In round 2, first avatar $A$ submits his physical identity parameters as a response and throws a challenge to avatar $B$, then $B$ submits his own physical identity parameters as a response and sends a session parameter to $A$. The specific steps are as follows:  (i) after $A$ checks the validity of $B$'s virtual identity based on $MIT_b$ and $VID_b$, $A$ submits $ PID_a$ and $C_b $ to $B$, where $C_b$ is the random challenge to $B$'s physical identity; (ii) upon $B$ checking  the consistency of $A$'s physical identities based on $MIT_a$ and  $PID_a$, $B$ responses $PID_b$ and $ g^w$ to $A$, among them, $g^w$ is the secure parameter for constructing session key $K=y^w$ between $A$ and $B$. 

\textbf{Session Key Establishment:}
After $A$ checking the validity of $B$'s physical identity based on  $MIT_b$ and $PID_b$, $A$ establishes the session key $K = (g^w)^{1/x} = y^w$ using his private key $sk=x$ to realize secure communication.

\section{Security Analysis} \label{sec:SecuAnaly}
 The security of the proposed authentication framework we constructed depends on the proposed collision chameleon signature. In this section, we first analyze the security of the chameleon collision signature, including the existential unforgeability under adaptive chosen message attacks (EUF-CMA) and the keys holding cost. Then, we analyze the security of the authentication framework, involving consistency, traceability, decentralized authentication, immersive mutual authentication,  key escrow problem, and privacy protection.

\subsection{Security of Chameleon Signature }

\textbf{EUF-CMA:}
The security of the proposed chameleon collision signature is based on  the Divisible computation Diffie-Hellman (DCDH) assumption\cite{Bao2003Variations}. 
Its security model is based on the existential unforgeability game $EUF^{DCDH}_\mathcal{A}(\mathcal{K})$ under chosen message attack. The game  $EUF^{DCDH}_\mathcal{A}(\mathcal{K})$ shown as Fig.\ref{fig:EUF} contains an adversary $\mathcal{A}$ and a challenger $\mathcal{B}$,  simulating the operation of a challenger and answering the queries from $\mathcal{A}$. 

\begin{figure}[htbp]
  \begin{center}
        \fbox{\parbox{8.2cm}{
        \begin{flushleft}
          \hspace{1em}$\textbf{Experiment} \quad EUF^{DCDH}_\mathcal{A}(\mathcal{K})$\\
          \hspace{1em}$parm \leftarrow Setup(\mathcal{K})$\\
          \hspace{1em}$(pk,sk) \leftarrow KeyGen(parm)$\\
          \hspace{1em}$(h,R) \leftarrow Hash(pk,M)$\\
          \hspace{1em}$\mathcal{O} \leftarrow \varnothing $\\
          \hspace{1em}$(M^*,R^*) \leftarrow \mathcal{A}^{H_r(\cdot),Sign^\prime(\cdot)}(pk,h,M,R) $\\
          \hspace{2em}where oracle $H_r, Sign^\prime$ on input $M^\prime, r^\prime$ respectively\\
          \hspace{3em}if $M^\prime \in \mathcal{O}$  return $\bot$\\
          \hspace{3em}$ r^\prime \leftarrow H_r(M^\prime)$\\
          \hspace{3em}$ R^\prime \leftarrow Sign^\prime(r^\prime) $\\
          \hspace{3em}$\mathcal{O} \leftarrow \mathcal{O} \bigcup \{M^\prime,R^\prime\}$\\
          \hspace{3em}return $R^\prime$ \\
          \hspace{1em}if $Verify(pk,h,M,R,M^{*},R^{*})=1\bigwedge (M^*,R^*)\notin \mathcal{O}$ \\
          \hspace{2em}return 1\\
          \hspace{1em}return 0
        \end{flushleft}
        }}

    \caption{The game model of the chameleon collision signature.}
    \label{fig:EUF}
    \end{center}
\end{figure}

In the following, we briefly analyze the security of the proposed chameleon signature algorithm, while the detailed proof is in the Appendix \ref{sec:AppProof}.

The security of the proposed chameleon signature algorithm depends on the DCDH assumption, that is, it is hard to output $g^{a/b} \in \mathbb{G}$ on given random triples $g, g^a, g^b \in \mathbb{G}$. According to the game model at Fig.\ref{fig:EUF}, the adversary attempts to forge a check parameter $R^*=(h/m^*)^x$ with respect to a new message $M^*$ through $(pk,h,M,R)$. Let $pk=g^{1/x}=g^A$, $(h/m^*)=g^B$, where $m^*=H(M^*)$, if the polytime adversary is able to output $R^*=( h/m^*)^x=g^{B/A}$, then the DCDH problem can be solved (it contradicts with the DCDH assumption). Therefore, the proposed chameleon signature is EUF-CMA.

\textbf{Keys Holding Cost:}
We compare the keys holding cost of our proposed chameleon signature with that of Khalili's  \cite{Khalili2019Efficient} and Gamenisch's \cite{Camenisch2017Trapdoors}.  From the \ref{sec:ChamSign} we get that the public and private key of our scheme contain one element in $\mathbb{G}_1$ and one element in $\mathbb{Z}_p$, respectively. For Khalili's scheme, it constructs two sets of chameleon hash methods in which construction 2 is more efficient, therefore, we only analyze it. Since the construction 2 leverages the signature of knowledge \cite{Jens2017Snarky} to ensure the security of the chameleon hash, the elements contained in the keys are greatly increased. By analyzing this construction, the elements contained in the key pair are shown in Table \ref{tab:Keys}. Its public key involves 3 group elements $( 2 \; G_1, 1 \; G_2 )$ and a set of common reference string $crs ( m+2n+5 \; G_1, n+3 \; G_2)$. Therefore,  Khalili's public key contains $ ( m+2n+7 \; G_1, n+4 \; G_2)$ and private key contains 3 parameters in $\mathbb{Z}_p$, which are $x\in \mathbb{Z}_p$ in chameleon hash and $ a, b \in \mathbb{Z}_p$  in zero-knowledge-proofs. For Gamenisch's scheme, it builds a chameleon hash algorithm in Gap-Groups,  which is the same as the group of our scheme. By analyzing this  construction, we get that its public key contains $(1\;G_1+2\;G_2 )$ and private key contains $2\;Z$. Thus, our scheme has low element cost in public-private key.  Additionally, Khalili's and Gamenisch's construction have to add a pair of public-private keys for signing because their keys can only generate a hash. But for our proposed algorithm, it only needs a pair of keys for hashing and signing. Therefore, our scheme has fewer key pairs to generate signatures. From the above analysis, we can get that our proposed chameleon signature has obvious advantages in terms of key holding cost under the premise of security.

\begin{table}[htbp]
\centering
\caption{ Elements of the Chameleon Key Pair}
\label{tab:Keys}
\begin{threeparttable}
    \begin{tabular}{p{50 pt}<{\centering}p{45 pt}<{\centering}p{45 pt}<{\centering}p{60 pt}<{\centering}}
        \hline
         \specialrule{0em}{1pt}{1pt}
        Construction & Hash PK \tnote{1} & Hash SK & Additional Key \tnote{2}\\
        \hline
        \specialrule{0em}{1pt}{1pt}
       \multirow{2}{*}{Khalili\cite{Khalili2019Efficient}} &$m+2n+7 \; G_1$ &\multirow{2}{*}{$3 \; Z $} &	\multirow{2}{*}{Yes}	\\
                                                           &  $ n+4 \; G_2  $ & &  \\
        \specialrule{0em}{1pt}{4pt}
       Camenisch\cite{Camenisch2017Trapdoors} &	$ 1\;G_1 + 2\;G_2 $ &	$ 2\;Z $ & Yes \\
        \specialrule{0em}{1pt}{1pt}
       This work &	$ 1 \; G_1  $ &	$ 1 \; Z  $  &	No  \\
        \hline
    \end{tabular}
\begin{tablenotes}
        \footnotesize
        \item[1] We utilize $G_1$, $G_2$, $Z$ to denote the number of elements contained in groups $\mathbb{G}_1$, $\mathbb{G}_2$, and finite field $\mathbb{Z}_p$, respectively.
        \item[2]  It indicates whether an additional key is required to implement the chameleon signature.
      \end{tablenotes}
\end{threeparttable}
\end{table}

\subsection{Security of Authentication Framework}

\emph{Deﬁnition 1:}
If an avatar's virtual identity and physical identity match a metaverse identity token, the avatar's identity satisfies consistency.

\textbf{Consistency:}
To guarantee the consistency, three aspects should be checked according to the \emph{Definition 1} :  the validity of the avatar's virtual identity, the validity of the avatar's physical identity, and the freshness of the submitted identity parameters. 
During an interaction between avatar $A$ as the prover and avatar $B$ as the verifier, (i) $B$ checks the match of $(M_a,R_a)=VID_a$  and $(M,R) \in MIT_a$ to ensure the validity of $A$'s virtual identity; (ii) $B$ checks the match of  $(M_a^\prime,R_a^\prime)=PID_a$ and $(M,R) \in MIT_a$ sampled from $A$'s manipulator to ensure the validity of $A$'s physical identity; (iii) $B$ first checks the match of $B$ and $T \in MIT$, then it extracts the watermarking $C_a^\prime$ from $M_a^\prime$ and compares $C_a^\prime$  with the challenge $C_a$ throwing by him to ensure that the biometric feature is freshness. To dynamically check the consistency of $A$'s virtual and physical identities, $B$ dynamically executes the one-party authentication protocol to have $A$ submit the biometric feature containing a new challenge. Through the above steps, it guarantees the consistency of the avatar's virtual and physical identities.

\textbf{Traceability:}
The virtual-physical traceability means that a malicious manipulator should be traced back through the avatar's identity parameters, a honest manipulator should not be framed by fake identity parameters. 

To track back a malicious manipulator, whistleblower submits IDP with the avatar's identity parameters $MIT=(T,y,h,M,R)$, $VID=(M_a,R_a)$  and $ PID =(M_a^\prime,R_a^\prime)$. After IDP completing the check on these parameters, it discloses the manipulator’s real identity based on $\{ID,MIT\}$ reserved registration, which achieves the virtual-physical tracking.  

Conversely, if a malicious whistleblower wants to frame an honest avatar,  the whistleblower needs to forge the avatar's $MIT$, $VID$, and $PID$. Among them, the $MIT$ and the $VID$ is the avatar's public information in the metaverse. Thus,  the target parameter whistleblower attempts to forge is the $PID$, that is, the  biometric feature $M_a^\prime$ and the corresponding check parameter $R_a^\prime $.  For $M_a^\prime$, the whistleblower is able to embed the fresh challenge $C_a$ into the outdated biometrics, which is easy to accomplish. But for $R_a^\prime$, since the whistleblower lacks the corresponding chameleon private key and the proposed chameleon signature is  EUF-CMA, it is hard for a whistleblower  to forge $R_a^\prime$.  Therefore, the proposed authentication framework satisfies the virtual-physical traceability.

\textbf{Decentralized Authentication:}
The decentralization authentication of the proposed framework relies on the MIT signed by IDP. In the process of authentication, the verifier checks the validity of the avatar's virtual identity based on $MIT$ and $VID$, checks the consistency of the avatar's virtual and physical identities based on $MIT$ and $PID$. The entire authentication process does not involve a trusted third party, realizing decentralized authentication between avatars.

\textbf{Immersive Mutual Authentication:}
 In the process of authentication,  the verifier throws a random challenge to the prover, then the prover automatically generates physical identity parameters as a response, such as capturing biometrics through a head-mounted display(HMD) devices, ensuring that the player's interactive experience is not disturbed in any way.

\textbf{No Key Escrow Problem:}
During authentication, the check on the avatar's virtual and physical identities is completed only by a pair of chameleon keys $(sk=x, pk=y)$, holding by the player himself. Therefore, the player's keys are not escrowed with any trusted third party, which avoids the key escrow problem.

\textbf{Privacy Protection:}
In the process of authentication between avatars,  the prover initiates an anonymous claim through $MIT$ and $VID$, at the same time, the verifier achieves anonymous authentication based on the locally processed biometric feature $M_a^\prime$ and the corresponding check parameter $R_a^\prime$. The whole process doesn't reveal the actual information about the prover's identity.

\section{Performance Evaluation}  \label{sec:PerfEval}
In this section, we first evaluate the computation cost of the chameleon signature. Then, we evaluate the performance of the proposed authentication framework, including the impact of embedding watermark into biometrics, the computational cost in authentication,  the authenticating consumption on different platforms, and the virtual-physical tracking consumption .

\subsection{Performance of Chameleon Signature}

\textbf{Computation Cost:}
We compare the computation cost of our proposed chameleon collision signature with that of Khalili's  \cite{Khalili2019Efficient} and Gamenisch's \cite{Camenisch2017Trapdoors} as shown in the TABLE \ref{tab:ComputationCost}. It can be seen that the proposed algorithm has obvious computational advantages. The reason is that Khalili's  and Camenisch's construction involves the signature of knowledge and zero-knowledge proof, respectively, while our algorithm eliminates the above time-consuming steps under the premise of security and greatly reduces the computational cost.

\begin{table}[htbp]
\centering
\caption{Computation Cost of Chameleon Signature}
\label{tab:ComputationCost}
\begin{threeparttable}
      \begin{tabular}{p{45 pt}<{\centering}p{60 pt}<{\centering}p{40 pt}<{\centering}p{60 pt}<{\centering}}
        \hline
        \specialrule{0em}{1pt}{1pt}
        Construction &  $Hash$ \tnote{1} & $Check$ & $Forge$ \tnote{2}\\
        \hline

        \specialrule{0em}{1pt}{1pt}
        \multirow{2}{*}{Khalili\cite{Khalili2019Efficient}}  & $ 3+m+2n-l \; E_1  $ & $l \; E_1$	& $2+m+2n-l \; E_1$ \\
         & $ n\;E_2+1\;M_1 $ & $ 4\;M_1+5 \; P $  & $n\;E_2+1\; M_1$ \\

        \specialrule{0em}{1pt}{4pt}
        \multirow{2}{*}{Camenisch\cite{Camenisch2017Trapdoors}} & $ 4\; E_1+4\; E_2  $ & $1\; E_1$	& $3\;E_1+1\;E_2$ \\
          & $ 1\;E_T+3\;P  $ & $ 1\;E_T+1\;P $  & $2\;E_T+2\;P$ \\

        \specialrule{0em}{1pt}{4pt}
        \multirow{2}{*}{This work} &	$2\;E_1$ &$1\;M_1$ &$1\;E_1$ \\
                              & $1\;M_1$ &$2\;P$ & $1\;M_1$ \\
        \specialrule{0em}{1pt}{1pt}
       \hline
      \end{tabular}
      \begin{tablenotes}
        \footnotesize
        \item[1] We utilize $E_1, E_2$, and $E_T$ to denote the exponential operation on $\mathbb{G}_1, \mathbb{G}_2$, and $\mathbb{G}_T$,  $M_1$ to denote the multiplication operation on the group $\mathbb{G}_1$, $P$ to denote the bilinear map. 
        \item[2] We utilize $Forge$ to denote the $HCol$ in  Khalili's construction, the $Adapt$ in Camenisch's construction,  and the $Sign$ in our scheme because the algorithms $HCol, Adapt$, and $Sign$ are essentially forging chameleon collisions.
      \end{tablenotes}
\end{threeparttable}
\end{table}

\begin{figure}[htbp]
\begin{center}
    \includegraphics[width=0.47\textwidth]{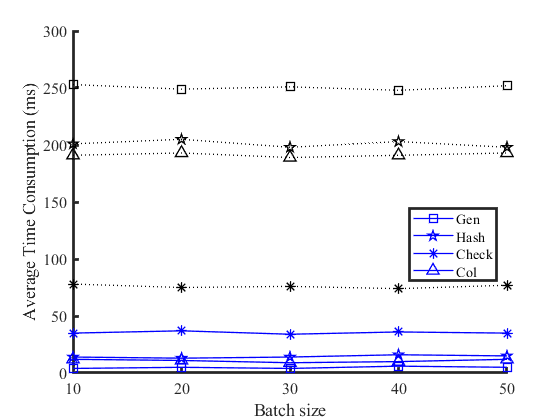}
    \caption{\small{The dotted line and the solid line in the figure are the time consumption of Khalili's  chameleon hash construction and our proposed chameleon signature algorithm, respectively.}}
    \label{fig:PropCham}	
 \end{center}
\end{figure}

To further analyze the performance of the proposed chameleon collision signature,  we implemented our scheme and Khalili's construction based on the Java programming language. The simulation set the batch size of data processing to $10, 20,30,40,$ and $50$ respectively. The average time consumption of these two methods is shown in Fig. \ref{fig:PropCham}. It can be seen from the figure that the consumption of our proposed algorithm is within 50ms, which has obvious advantages compared with Khalili's construction.

\subsection{Performance of Authentication Framework}
Since the metaverse platform presents vivid digital environments via HMD devices that fully cover the player's eye area to guarantee a perfect visual experience, we consider the iris-based authentication method will be an important way to achieve metaverse authentication.  Based on this idea, the proposed protocol embeds challenge information into the iris feature to realize the avatar authentication and prevent replay attacks. 

\begin{figure}[htbp]
\begin{center}
    \includegraphics[width=0.47\textwidth]{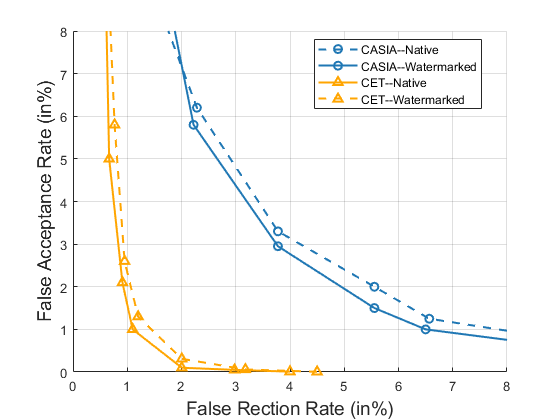}
    \caption{\small{the FRR and the FAR of iris recognition.}}
    \label{fig:FrrFar}
 \end{center}
\end{figure}

\textbf{Watermarked Biometrics:}
To analyze the impact of embedding watermarks into biometrics, we take the player's native iris feature as the biotemplate $T$ in the avatar's $ MIT$,  take the native and the watermarked iris features as the response parameters, respectively. During simulation, we utilize the public dataset CASIA-IrisV4-Thousand \footnote{http://biometrics.idealtest.org} and CET2005 \footnote{https://www.nist.gov/itl/iad} to build the iris biometrics, the OSIRIS algorithm \cite{Othman2016OSIRIS} to extract the iris-encoded, and the adaptive image watermarking algorithm \cite{Huang2019DCT2} to embed 128-bit random challenge into the iris-encoded image. Through simulations, we obtain the False Rejection Rate (FRR) and the False Acceptance Rate (FAR) of iris recognition as shown in \ref{fig:FrrFar}. It can be seen that although the proposed authentication framework embeds a watermark into the iris-encoded, the FRR and FAR of iris recognition are almost the same as that of the native iris features under the two types of public datasets. Therefore, our method further prevents replay attacks without affecting the recognition efficiency.

\textbf{Computational Cost of the Protocol:}
The proposed authentication protocol contains one-party and two-parties. Since the latter includes all the steps of the former, we analyze the computational cost of the latter. The two-parties protocol involves verifying the signature on $MIT$, embedding and extracting watermarks in biometric features, and generating and checking chameleon parameters. To analyze the core computational cost of our proposed protocol, we only consider the generating and checking of chameleon parameters. From the two-parties protocol in \ref{sec:TwoPartAuth}, the costs of each phrase are shown in the TABLE \ref{tab:ProtocolCost}. 

In round 1, avatar $A$ only submits the identity parameter to $B$, so no computation is involved,  while the avatar $B$ need to check the validity of avatar $A$'s identity through $Verify$, which involves the computational cost of $(1\;M_1+2\;P)\times2$. In round 2, $A$ need to check the validity of $B$'s identity through $Verify$ and generate a chameleon signature by $Sign$, which involves the calculation cost of $(1\;M_1+2\;P)\times2+1\;E_1+1\;M_1$.  $B$ need to check the validity of $A$'s response by $Verify$, generate a chameleon signature by $Sign$, and submit a random parameter $g^w$ to $A$, which involves the computation cost of $(1\;M_1+2\;P)\times2+2\;E_1+ 1\;M_1$. In session key establishment, $A$ need to check the validity of $B$'s response through $Verify$ and generate the session key $K=(g^w)^{1/x}$ by his chameleon private key $x$, which involves the computation cost of $(1\;M_1+2\;P)\times2+1\;E_1$. But for $B$, he establishes the session key $K=y^w$ by the random number $w$, which involves the computational cost of $1\;E_1$. To sum up, the calculation costs of $A$ and $B$ in the two-parties authentication protocol are $(5\;M_1+2\;E_1+8\;P)$ and $(5\;M_1+3\;E_1+8\;P)$, respectively.

\begin{table}[htbp]
      \centering
      \caption{Computation Cost of Two-Parties Authentication Protocol}
      \label{tab:ProtocolCost}
      \begin{tabular}{p{40 pt}<{\centering}p{88 pt}<{\centering}p{88 pt}<{\centering}}
        \hline
        \specialrule{0em}{1pt}{1pt}
        Phase & avatar $A$ & avatar $B$\\
        \hline
        \specialrule{0em}{1pt}{1pt}
        Round 1 & -- & $2\;M_1+4\;P$  \\
        \specialrule{0em}{1pt}{1pt}
        Round 2 & $3\;M_1+4\;P+1\;E_1$   & $ 3 \; M_1 + 4 \; P + 2 \; E_1$  \\
        \specialrule{0em}{1pt}{1pt}
        Session  & $2 \; M_1 + 4 \; P + 1\; E_1$ & $  1 \; E_1  $  \\
        \specialrule{0em}{1pt}{1pt}
        Total & $5\;M_1+8\;P+2\;E_1$ & $5\;M_1+8\;P+3\;E_1$ \\
        \hline
      \end{tabular}

\end{table}

\begin{figure}[htbp]
\begin{center}
    \includegraphics[width=0.47\textwidth]{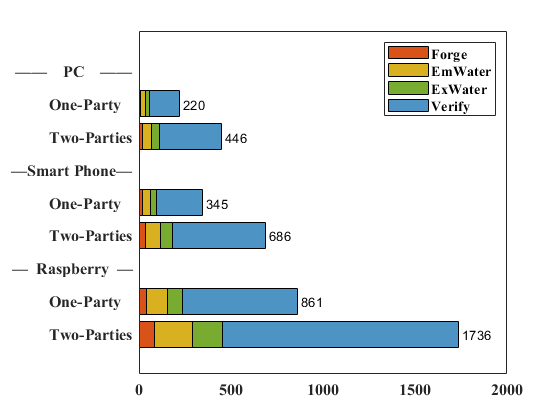}
    \caption{\small{ Authentication consumption on different platforms.}}
    \label{fig:PlatformsTime}
 \end{center}
\end{figure}

\textbf{Authenticating Consumption on Different Platforms:}
To further analyze the time consumption of the proposed authentication framework, we  conduct simulations on three different platforms, PC ( CPU is Intel Core i5 7500, memory is 8 GB, the operating system is Windows 10 ), Smart Phone(CPU is Hisilicon Kirin 659, memory is 4 GB, the operating system is Android 9), and Raspberry(CPU is Cortex-A72, memory is 4 GB). The time consumption of one-party authentication protocol and two-parties authentication protocol on different platforms are shown as Fig.\ref{fig:PlatformsTime}, where $Forge$ is the time consumption to generate the chameleon signature, $EmWater$ is to embed random challenge as a watermark into the biometric feature, $DeWater$ is to extract the watermark from the biometric feature, and $Verify$ is to check the validity of avatar's virtual and physical identities. It can be seen from Fig. \ref{fig:PlatformsTime} that the total consumption of the two-parties authentication protocol on PC and Smart Phone is within 1000ms, while that on Raspberry is about 2000ms. In general, the proposed authentication framework is efficient on PC and Smart Phone.

\begin{figure}[htbp]
\begin{center}
    \includegraphics[width=0.47\textwidth]{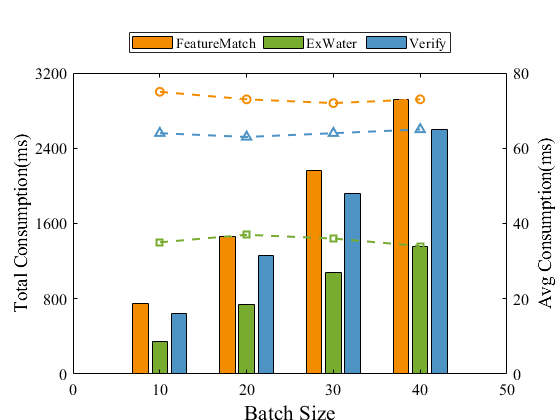}
    \caption{\small{The consumption of virtual-physical tracking on core steps.}}
    \label{fig:TrackTime}
 \end{center}
\end{figure}

\textbf{Virtual-physical Tracking Consumption on Core Steps:}
The identity parameters retained in the authentication process provide conditions for virtual-physical tracking. The specific steps are as follows: (i) the whistleblower submits IDP with the avatar's identity parameters including $MIT,VID$, and $PID$; (ii) the IDP  checks the validity of virtual identity based on $MIT$ and $VID$, checks the consistency of physical-virtual identity based on $MIT$ and $PID$; (iii) after the above check is passed, the IDP reveals the manipulator's real identity via $\{ID,MIT\}$, which achieves virtual-physical tracking. The time consumption of the above process mainly involves the feature matching between the iris-encode and biometric template,  the extracting watermark from iris-encode, and the verifying signature on the chameleon collision pair. To analyze this consumption, we set the report times for malicious avatars to be 10, 20, 30, and 40 respectively. Through simulations, we get the tracking consumption for virtual-physical tracking as shown in Fig.\ref{fig:TrackTime}. What can be seen from the figure that the average consumption of feature matching is about 75ms, the extracting watermark is about 35ms, and the verifying signature is about 65ms. From the above cost analysis, we can get that the designed authentication framework can not only achieve virtual-physical traceability but also has a low consumption cost.

\section{Related work} \label{sec:RelaWork}
Metaverse players manipulate their avatars based on head-mounted display devices. In building an authentication framework, if the avatar is checked only by the device keys, it is hard to ensure the consistency between the avatar in virtual space and its manipulator in physical world; if the avatar is checked only by the player's biometrics, it is difficult to avoid biometrics replay. Therefore, combining device keys and player biometrics is critical for building a traceable authentication framework. At present, researchers mainly design authentication protocols for users or devices through two types of methods, knowledge-based and biometric-based. To the best of our knowledge, there is no authentication mechanism for avatars in the metaverse. In this section, we sort out the above authentication methods to build a traceable authentication framework for avatars.

\begin{table*}[htbp]

      \centering
      \caption{Comparison of the advantages of different solutions}
      \label{tab:Advantages}
      \begin{threeparttable}
      \begin{tabular}{p{90 pt}<{\centering}p{70 pt}<{\centering}p{45 pt}<{\centering}p{45 pt}<{\centering}p{45 pt}<{\centering}p{45 pt}<{\centering}p{45 pt}<{\centering}p{45 pt}<{\centering}}
        \hline
        \specialrule{0em}{1pt}{1pt}
        Category & Factors & $Virtual$  & $Physical$  & $Immersive$ & $Continuous$  & $Mutual$  & $Session Key$ \\
        \hline 
        \specialrule{0em}{1pt}{1pt}
        Knowledge & Password \cite{Szalachowski2021Password} & \ding{52}  & \ding{56}  & \ding{56} & \ding{56} & \ding{52}  & \ding{52}\\
        Knowledge & Key \cite{Gope2019Lightweight} &  \ding{56} & \ding{52} & \ding{52}  & \ding{52} & \ding{52}  & \ding{52}\\
        Biometric& Physiology \cite{Luo2020OcuLock} & \ding{52} & \ding{52} & \ding{52} & \ding{56} & \ding{56}  & \ding{56}\\
        Biometric& Behavior \cite{Lee2021Usable} &  \ding{52} &  \ding{52} & \ding{52} & \ding{52} & \ding{56}  & \ding{56}\\
        Knowledge-Biometric & Multi-Factor \cite{Gunas2018PrivBi} &   \ding{52} &  \ding{52}  &  \ding{56} &  \ding{56} &  \ding{52}  &  \ding{52}\\
        \hline
      \end{tabular}
      \begin{tablenotes}
        \footnotesize
        \item[] We utilize $Virtual$ and $Physical$ to denote the verifiability of an entity's virtual identity and physical identity respectively, $Immersive$ indicates the immersive authentication between entities, $Continuous$ represents the continuous authentication, $Mutual$ as the mutual authentication, and $Session Key$ as the negotiation of session keys.
      \end{tablenotes}
\end{threeparttable}
\end{table*}

\subsection{Knowledge-Based Authentication}

Knowledge-based authentication methods discriminate the identities of users or devices based on information known only to the entity. For example, the network platform authenticates the login user via their password, and the Internet of Things authenticates the device based on its key.

\textbf{Password-authenticated:} The most widespread authentication method over the Internet is the password-authenticated mechanism. The mechanism relies on server operators as trusted parties to offer secure authentication, giving them full control over users' identities. Zhang \cite{Zhang2021Threshold} constructed a password-based threshold single-sign-on authentication scheme to thwart adversaries from compromising the identity servers. To mitigate the threat of centralized authentication, Szalachowski \cite{Szalachowski2021Password} designed a password-authenticated decentralized identityn by which users register their self-sovereign username-password pairs and use them as universal credentials.  This kind of password-authenticated method can realize the security authentication between the device and the service provider. However, these methods require the user to enter a password during the authentication process, which fails to achieve immersive authentication. In addition, anyone can pass the login authentication as long as he knows the password, making this method impossible to guarantee the consistency between the avatar and its manipulator.

\textbf{Key-authenticated:} With the large-scale application of IoT devices, researchers have been working on building a secure and efficient authentication framework\cite{Cui2020Hybrid}  based on public-private key pairs. Guo \cite{Guo2020Blockchain} designed an asymmetric cryptographic algorithm based on elliptic curve cryptography (ECC) to construct a distributed and trusted authentication system for smart terminals. To achieve that the users' identifying information is controlled by himself, Xu \cite{Xu2020Identity} constructed blockchain-based identity management and authentication scheme. Shen \cite{Shen2020Secure} presented an efﬁcient blockchain-assisted secure device authentication mechanism for cross-domain IIoT to alleviate the cost of key management overhead and eliminate the problem of relying on a trusted third party. These public key-based authentication methods facilitate one-party authentication without involving a trusted third party, but it is difficult to efficiently realize Device-to-Device (D2D) mutual authentication. Certificateless cryptographic algorithms are widely used to build mutual authentication methods between devices \cite{Wang2019Auten} because it avoid key escrow issues.  Shang \cite{Shang2020Secure} designed an authentication and key agreement protocol to negotiate a group session key securely and effectively in D2D group communications, which merges the advantages of certificates public key cryptography (CL-PKC) and ECC.  Gope \cite{Gope2019Lightweight} presents a lightweight and privacy-preserving two-factor authentication scheme in which physically unclonable functions(PUF) have been considered as one of the authentication factors. Li \cite{Li2021Provably} proposes an end-to-end mutual authentication and key exchange protocol for IoT by combining PUF with CL-PKC on an elliptic curve. The above scheme realizes efficient mutual authentication between devices, but the authentication process is separated from the user. If this kind of method is directly used to build the metaverse authentication framework, it is also difficult to ensure the consistency of an avatar and its manipulator. Additionally, the process of generating signatures is cumbersome, which imposes huge computational and communication costs. Therefore,  we desire to construct an efficient signature algorithm based on the ideas of chameleon hash \cite{Khalili2019Efficient} and short signature \cite{Boneh2004Short} to realize the immersive authentication for the avatar's virtual identity.

\subsection{Biometric-based authentication}
Biometrics are inherent information of the human body and there is no risk of being forgotten or lost, making these methods widely used in the verification of the user's physical identity. The type of biometric-based authentication method can be divided into two sub-categories, physiological biometric-based and behavioral biometric-based. The physiological biometric-based methods are primarily based on face \cite{Luo2021Face}, iris \cite{Bernadelli2021Iris}, fingerprint \cite{Chen2022Fingerprint} and other characteristics to complete static authentication. The behavioral biometric-based methods rely on walking gait \cite{Zhang2022Gait}, eye movements \cite{Zhang2018EyeMove}, and response to vibrations \cite{Lee2021Usable} to achieve dynamic authentication.

\textbf{Physiological biometric-based:} Owing to its simplicity and accuracy, physiological-based authentication technology has attracted extensive attention of many experts and scholars \cite{Luo2020OcuLock} \cite{Agrawal2020Mobile} \cite{Perera2018Multiple}. Papadamou \cite{Papadamou2018Killing} proposed a privacy-preserving federated architecture for device-centric authentication (DCA) in which the core authentication functionality resides on a trusted entity. To prevent the attacker from knowing the user's biometrics after the database is compromised, Chatterjee \cite{Chatterjee2019Multisket} proposed practical secure sketches to conceal the correspondence between users and their biometric templates. The above methods realize efficient authentication of device and user identities, but these methods are all one-time authentication methods, which are difficult to guarantee that the user's identity is consistent with avatar's identity.

\textbf{Behavioral Biometrics-based:} To mitigate the vulnerabilities of static biometrics, the behavior-based authentication method realizes dynamic authentication by continuously collecting user biometrics. Zhu \cite{Zhu2020RiskCog} designed an unobtrusive real-time user authentication system by which each biometric sample is aggregated into users' implicit events, such as raising hands to check the time on the watch. Li \cite{Li2019Velody} leverages the response of hand-surface vibration to construct a resilient user authentication system. Similarly, Lee \cite{Lee2021Usable} proposed a usable method for user authentication through the response to vibration challenges on the user's smartwatches. Wu \cite{Wu2020LVID} designed a multimodal biometrics system on smartphones that leverages lip movements and voice for authentication. It can be seen from the above scheme that the behavioral-mentioned authentication mainly solves the problem of dynamic authentication without involving mutual authentication between users. Moreover, these methods are beneficial to complete the continuous authentication between the user's physical identity and their device but are easily limited by physical scenarios, such as, the vibration-response method may be invalid when the user has some items in their hands because the newly added item may interfere with the response value\cite{Liang2020Continuous}.

Multi-factor authentication mechanism is conducive to building an authentication mechanism that meets any scenario, as it eliminates dependence on  trusted third parties and avoids the involvement of  user behavior.  Gunasinghe \cite{Gunas2018PrivBi} introduces a three-factor authentication scheme based on a signed identity token that encodes the user's biometric identifier and the password entered by the user into the token. Although this method requires users to enter a password during the authentication process, which disturbs the immersive experience of metaverse players, this method is facilitative to the realization of decentralized mutual authentication for avatars based on player's biometrics and his keys. Inspired by this idea, we designed a decentralized authentication framework based on two factors to achieve the verifiability on avatar's virtual and physical identities and guarantee the virtual-physical traceability.

\section{Conclusion} \label{sec:Conclu}
Metaverse is a virtual-physical coexistence environment where people communicate and work via digital avatars. In the emerging social ecosystems, however, malicious players frequently violate the safety of other avatars, posing a huge challenge to the healthy development of metaverse. For this issue, we design a two-factor authentication framework based on chameleon signature and biometrics, which guarantees the virtual-physical traceability that tracking an avatar in virtual space to its manipulator in physical world. To the best of our knowledge, our method is the first work for avatars tracking in the metaverse field. We hope that the proposed framework could bring a little reference to researchers in related fields.

\appendices
\section{Security Proof for Chameleon Collision Signature}\label{sec:AppProof}

\textbf{Theorem:} Let $\mathbb{G}$ be a multiplicative cyclic  group and $H_\mathbb{G}$ be a collision-resistant hash on $\mathbb{G}$, if the DCDH assumption holds on $\mathbb{G}$, the chameleon collision signature is EUF-CMA.

\textbf{Proof:}  suppose there is a polytime adversary $\mathcal{A}$ that breaks the chameleon collision signature with the advantage of $\epsilon(\mathcal{K})$, then there must be an adversary $\mathcal{B}$ to solve the DCDH on $\mathbb{G}$ at least by the advantage of
 $$Adv^{DCDH}_\mathcal{B} \geq \frac{\epsilon(\mathcal{K})}{e \cdot q_H}.$$

Where $e$ is the base of the natural logarithm, $q_H$ is the maximum number of queries to $H_\mathbb{G}$.

From section \ref{sec:SecuAnaly}, the process of the game $EUF^{DCDH}_\mathcal{A}(\mathcal{K})$ between $\mathcal{A}$ and $\mathcal{B}$ is as follows:

(1) Adversary $\mathcal{B}$ runs  $Setup(\mathcal{K})$ and $KeyGen(Parm)$ to select a random function  $H_\mathbb{G}\stackrel{R}{\leftarrow}\{ H:\{0,1\}^*\rightarrow \mathbb{G}\}$, generate the key pair $(pk,sk)$, calculate the original chameleon parameters $( y, h, M, R )$, and send adversary $\mathcal{A}$ with the system parameters and the chameleon parameters.

(2) The adversary $\mathcal{A}$ queries the adversary $\mathcal{B}$ for the hash and the corresponding signature with any $M^\prime$. Adversary B response $r^\prime$ and $R^\prime$ as the corresponding answer. During this process, $q_H$ is the maximum number of times $\mathcal{A}$ queries $H_\mathbb{G}(\cdot)$.

(3) The adversary $\mathcal{A}$ outputs a chameleon collision $(M^*, R^*)$ as a forged signature. If $Check(y,h,M,R) = Check(y,h,M^*,R^*)=1$, the adversary's attack is successful, where the check parameter $R^*$ of $M^*$  has not been queried by it before.

From the above process, the adversary $\mathcal{A}$ wants to find a certain $r^*$ related to $M^*$ such that $(r^*)^x=(h/m^*)^x=R^*$, then, a fake collision $(M^*, R^*)$ can be successfully output. During the hash query, if $m^*$ is the hash value of a certain message $M^*$, then $(h/m^*)^x = R^*$ is the check parameter for $M^*$. In step (3), $(M^*,R^*)$ is generated by adversary $\mathcal{A}$, but $H_\mathbb{G}(M^*)$ is generated by $\mathcal{B}$, so $\mathcal{B}$ is able to set $r^*=h/H_\mathbb{G}(M^*)=H_r(M^*)$. When $\mathcal{B}$ lets $r^*$ be the potential hash of a target message, his goal is to call $\mathcal{A}$ to calculate $(r^*)^x$ based on the triple $(g,g^{1/x},r^*)$, which is to solve the DCDH problem. As already proven in \cite{Bao2003Variations}, the DCDH problem is equivalent to the CDH problem. Thus, we convert $B$'s attack target to the CDH problem, that is, calling $\mathcal{A}$ to calculate $(r^*)^x$ based on the triple $(g,g^x,r^*)$. Throughout the game, $\mathcal{B}$ doesn't know which message will be generated by $\mathcal{A}$ to forge a check parameter. Therefore,  $\mathcal{B}$  has to make a guess that the \emph{j}-th query $H_r$ corresponds to  the final forged result from $\mathcal{A}$.

For simplicity without loss of generality, we assume that: (i). Adversary $\mathcal{A}$ will not initiate the same query $H_r(\cdot)$ twice to $M^\prime$; (ii). Adversary $\mathcal{A}$ must have asked $H_r(M^\prime)$ before querying the check parameter $R^\prime$; (iii). Adversary $\mathcal{A}$ must have asked $H_r(M^*)$ before he outputs $(M^*,R^*)$.

In the actual process, $\mathcal{B}$ implicitly regards $u=g^a$ in the known tuple $(g,u=g^a,r^*)$ as its own public key (in fact, $\mathcal{B}$ doesn't know the specific value of $a$, then $(r^*)^a$ is a forged check parameter of a certain message, namely $R^{*}=(r^*)^a=(h/H_\mathbb{G}(M^*))^a=(H_r(M^*))^a$, where $(r^*)^a$ is forged by $\mathcal{B}$. To hide instance $u=g^a$, $\mathcal{B}$ needs to select a randomness $t\stackrel{R}{\leftarrow} Z_p^*$ and send $u \cdot g^t$ to $\mathcal{A}$ as the public key of $\mathcal{B}$.

The following proves that the EUF-CMA game $EUF^{DCDH}_\mathcal{A}(\mathcal{K})$ of the chameleon collision signature can be reduced to the CDH problem.

(1) $\mathcal{B}$ sends the generator $g$ of group $\mathbb{G}$ and the public key $y$ to $\mathcal{A}$, where the private key corresponding to $y$ is $a+t$ , $t\stackrel{R}{\leftarrow} Z^*_q$. That is

$$y=u\cdot g^t=g^{a+t}\in \mathbb{G}.$$

 At the same time, $\mathcal{B}$ randomly selects $j\stackrel{R}{\leftarrow}\{1,2,\cdot\cdot\cdot,q_H\}$ as the hypothetical index of the forged parameters, that is, the \emph{j}-th query of $H_r$ from $\mathcal{A}$ corresponds to the hash of the target message $M^*$.

(2) $H_r$ query (at most $q_H$ times). $\mathcal{B}$ creates an empty list $H^{list}$ and lets the five-tuple $(h,M_i,r_i,b_i,r_i^\prime)$ be the element in it, which means that $\mathcal{B}$ has set

$$H_r(M^\prime)=h/H_\mathbb{G}(M_i)=r_i.$$

When $\mathcal{A}$ makes the \emph{i}-th inquiry to $H_r(\cdot)$, $\mathcal{B}$ randomly selects $b_i \stackrel{R}{\leftarrow} Z^*_p$ and answers as follows:

$\cdot$ If $i=j$, return $r_i^\prime=r_i\cdot g^{b_i}\in\mathbb{G}$;

$\cdot$ Otherwise, calculate $r_i^\prime=g^{b_i}\in\mathbb{G}$.

$\mathcal{B}$ takes $r_i^\prime$ as the answer to the query $H_r(M^\prime)$ and appends $(h,M_i,r_i,b_i,r_i^\prime)$ to the list $H^{list}$.

(3) $Sign^\prime$ query (at most $q_H$ times). In the process of $\mathcal{A}$ requesting the check parameter of a message $M^\prime$, $\mathcal{B}$ lets $M^\prime=M_i$ be the \emph{i}-th $H_r$ query, and answers the query in the following way:

$\cdot$ If $i \neq j$, $\mathcal{B}$ retrieves the tuple $(h,M_i,r_i,b_i,r_i^\prime)$ in the $H^{list}$ by $M_i$,  computes $R^\prime_i=(u\cdot g^t)^{b_i}$, and returns $R^\prime_i$ to $\mathcal{A}$, where $R^\prime_i$ is the check parameter constructed by ($r_i^\prime$, $M_i$) with secret key $a+t$. For 
$$R^\prime_i=(u\cdot g^t)^{b_i}=(g^{a+t})^{b_i}=(g^{b_i})^{a+t}=(r_i^\prime)^{a+t}.$$

$\cdot$ Otherwise, interrupt.

(4) $\mathcal{A}$ outputs $(M^{*},R^{*})$. If $M^{*} \neq M_j$, interrupt; otherwise, $\mathcal{B}$ outputs $\frac{R^{*}}{r^tu^{b_j}g^{b_jt}}$ as $(r^*)^a$. For
\begin{align}
\notag
R^{*}&=(r_j^\prime)^{(a+t)}=(r^* \cdot g^{b_j})^{(a+t)}\\
\notag
&= (r^*)^{(a+t)}\cdot g^{b_j(a+t)}\\
\notag
&=(r^*)^a r^t\cdot (g^{(a+t)})^{b_j}\\
\notag
&=(r^*)^a r^t\cdot u^{b_j}g^{b_jt}.
\notag
\end{align}
From this, we can get the following result:
$$\frac{R^{*}}{r^tu^{b_j}g^{b_jt}} = \frac{(r^*)^a r^t u^{b_j}g^{b_jt}}{r^tu^{b_j}g^{b_jt}}=(r^*)^a.$$

If the guess \emph{j} from $\mathcal{B}$ is correct and $\mathcal{A}$ finds a correct forgery, $\mathcal{B}$ successfully solves the given CDH problem, that is, $\mathcal{B}$ finds $R^{*}=(r^*)^a$ based on $(g,g^a,r^*)$ through $\mathcal{A}$. The successful output $R^{*}=(r^*)^a$ from $\mathcal{B}$ is determined by the following three events:

$\mathcal{E}_1$: No interruption was encountered during the interaction between  $\mathcal{A}$ and  $\mathcal{B}$.

$\mathcal{E}_2$: $\mathcal{A}$ produces a valid “message-parameter” pair $(M^{*},R^{*})$.

$\mathcal{E}_3$: $\mathcal{E}_2$ occurs and the subscript of $M^{*}$ in the corresponding five-tuple $(h,M_i,r_i,b_i,r_i^\prime)$ is $i=j$. Then

$Pr[\mathcal{E}_1]=(1-\frac{1}{q_H})^{q_H}$,

$Pr[\mathcal{E}_2|\mathcal{E}_1]=\epsilon(\mathcal{K})$,

$Pr[\mathcal{E}_3|\mathcal{E}_1\mathcal{E}_2] =Pr[i=j|\mathcal{E}_1\mathcal{E}_2] =\frac{1}{q_H}$ .

So the advantage of $\mathcal{A}$ is :
\begin{align}
\notag
Pr[\mathcal{E}_1\mathcal{E}_3]&=Pr[\mathcal{E}_1] \cdot Pr[\mathcal{E}_2|\mathcal{E}_1]\cdot Pr[\mathcal{E}_3|\mathcal{E}_1\mathcal{E}_2]\\
\notag
&=(1-\frac{1}{q_H})^{q_H}\cdot \frac{1}{q_H}\cdot \epsilon(\mathcal{K})\\
\notag
&\approx \frac{\epsilon(\mathcal{K}) }{e\cdot q_H}.
\end{align}

Since the DCDH assumption holds on $\mathbb{G}$ and the DCDH problem
equals to the CDH problem, the advantage $\epsilon(\mathcal{K})/(e\cdot q_H)$ of polytime adversary $\mathcal{B}$ is negligible, so the chameleon sign algorithm is EUF-CMA. (Theorem is proved)

\ifCLASSOPTIONcaptionsoff
  \newpage
\fi

\begin{IEEEbiography}[{\includegraphics[width=1in,height=1.25in,clip,keepaspectratio]{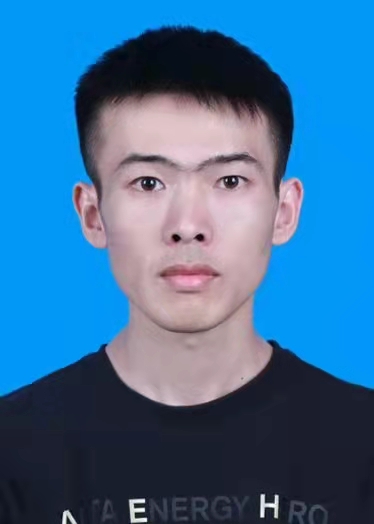}}]{Kedi Yang}
received the B.Sc. degree in mathematics and applied mathematics from Anshun University in 2012, and the M.Sc. degree in applied mathematics from Guizhou University in 2020.
He is currently a Ph.D candidate in the College of Computer Science and Technology, Guizhou University, Guiyang, China. His research interests mainly focus on Metaverse security, data provenance, and blockchain technology.
\end{IEEEbiography}

\begin{IEEEbiography}[{\includegraphics[width=1in,height=1.25in,clip,keepaspectratio]{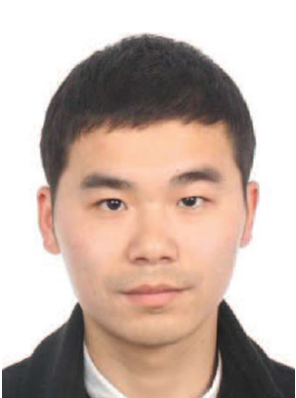}}]{Zhenyong Zhang}
 (Member, IEEE) received the achelor’s degree from Central South University, Changsha, China, in 2015, and the Ph.D. degree from Zhejiang University, Hangzhou, China, in 2020. He was a Visiting Scholar with Singapore University of Technology and Design, Singapore, from 2018 to 2019.  He is currently a Professor with the College of Computer Science and Technology, Guizhou University, Guiyang, China. His research interests include cyber–physical system security, applied cryptography, metaverse security, and machine learning security.
\end{IEEEbiography}

\begin{IEEEbiography}[{\includegraphics[width=1in,height=1.25in,clip,keepaspectratio]{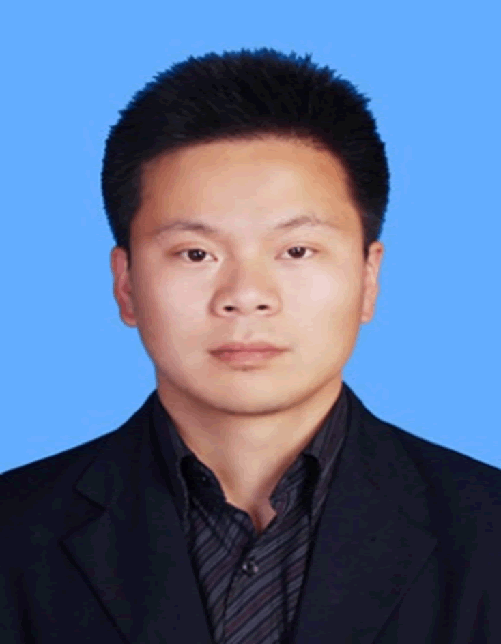}}]{Youliang Tian}
 (Member, IEEE) received the B.Sc. degree in mathematics and applied mathematics and the M.Sc. degree in applied mathematics from Guizhou University, in 2004 and 2009, respectively, and the Ph.D. degree in cryptography from Xidian University, in 2012. From 2012 to 2015, he was a Postdoctoral Associate with the State Key Laboratory for Chinese Academy of Sciences. He is currently a Professor and a Ph.D. Supervisor with the College of Computer Science and Technology, Guizhou University. His research interests include algorithm game theory, cryptography, and security protocol.
\end{IEEEbiography}

\begin{IEEEbiography}[{\includegraphics[width=1in,height=1.25in,clip,keepaspectratio]{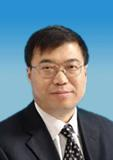}}]{Jianfeng Ma}
 (Member, IEEE) received the B.S. degree in mathematics from Shaanxi Normal University, Xi'an, China, in 1985, and the M.S. degree and the Ph.D. degree in computer software and telecommunication engineering from Xidian University, Xi'an, China, in 1988 and 1995, respectively. He is currently a professor with the School of Cyber Engineering, Xidian University, Xi'an, China. He is also the Director of the Shaanxi Key Laboratory of Network and System Security. His current research interests include information and network security and mobile computing systems.
\end{IEEEbiography}

\end{document}